\documentclass[aps, pra, showpacs, superscriptaddress, twocolumn, 10pt]{revtex4-1}

\usepackage{amsfonts,amssymb,amsmath,amsthm,longtable,array}
\usepackage{graphics,graphicx,color,enumitem}
\usepackage{epstopdf,epsfig}

\renewcommand{\>}{\rangle}
\newcommand{\be}{\begin{equation}}
\newcommand{\ee}{\end{equation}}
\newcommand{\ben}{\begin{eqnarray}}
\newcommand{\een}{\end{eqnarray}}

\begin{document}
	
	\title{Higher dimensional entanglement without correlations}
	
	\author{Waldemar K{\l}obus}
	
	\affiliation{Institute of Theoretical Physics and Astrophysics, Faculty of Mathematics, Physics and Informatics, University of Gda\'nsk, 80-308 Gda\'nsk, Poland}
	
	\author{Wies{\l}aw Laskowski}
	
	\affiliation{Institute of Theoretical Physics and Astrophysics, Faculty of Mathematics, Physics and Informatics, University of Gda\'nsk, 80-308 Gda\'nsk, Poland}
	
	\author{Tomasz Paterek}
	
	\affiliation{School of Physical and Mathematical Sciences, Nanyang Technological University, 637371 Singapore}
	
	\affiliation{MajuLab, CNRS-UCA-SU-NUS-NTU International Joint Research Unit, UMI 3654, 117543 Singapore}
	
	\author{Marcin Wie\'sniak}
	
	\affiliation{Institute of Informatics, Faculty of Mathematics, Physics and Informatics,
		University of Gda\'nsk, 80-308 Gda\'nsk, Poland}
	
	\author{Harald Weinfurter}
	
	\affiliation{Max-Planck-Institut f\"ur Quantenoptik, Hans-Kopfermann-Stra{\ss}e 1, 85748 Garching, Germany}
	
	\affiliation{Department f\"ur Physik, Ludwig-Maximilians-Universit\"at, 80797 M\"unchen, Germany}

	\begin{abstract}
		It has been demonstrated both theoretically and experimentally that genuine multipartite entanglement between qubits can exist even in the absence of multipartite correlations.
		Here we provide first examples of this effect in higher dimensional systems -- qudits.
		We construct states in which genuine $N$-partite entanglement between qudits is supported only by correlations involving strictly less than $N$ particles.
		The construction differs in several aspects from the ones for qubits.
		The states introduced here are a natural test-bed for candidate quantifiers of genuinely multipartite quantum correlations.
	\end{abstract}

	\maketitle
\section{Introduction}
\label{intro}

Quantum entanglement in pure states is typically presented as a situation where the whole system has smaller entropy than its subsystems.
There is, however, an alternative characterisation of entanglement.
Instead of focusing on subsystems, it turns out that a pure $N$-party state is entangled if and only if the sum of all its $N$-party correlations squared exceeds certain bound~\cite{HJ1,HJ2,HJ3,us1,us2}.
This holds for arbitrary number of particles and arbitrary dimensions.
One then wonders if a similar statement could be derived for mixed states,
i.e. whether there exists some processing of the correlations between all the particles (and only such correlations) which witnesses entanglement in every entangled mixed state.

It has been demonstrated both in theory~\cite{Dag,nc2,nc3} and in experiment~\cite{nc1} that in the case of many qubits such a procedure does not exist.
On one hand, clearly there are non-entangled states with vanishing correlations between all the parties, e.g. white noise.
On the other hand, there exist genuinely $N$-party entangled mixed states which are also endowed with vanishing correlations between all the observers~\cite{Dag,nc2,nc3,nc1}.
In principle, there still exists a possibility that a similar argument cannot be put forward for higher dimensions,
but here we show by explicit construction that mixed-state entanglement of qudits also cannot be witnessed by $N$-party correlations only.

The genuinely multiparty entangled states without multipartite correlations were originally introduced in the context of quantifying genuinely multiparty classical and quantum correlations and the states we introduce here are also valuable for this task.
Since up to date there are no computable quantifiers of genuinely multipartite classical or quantum correlations that satisfy natural postulates of Ref.~\cite{postulates}, these states are a nontrivial test-bed for the candidate such quantifiers.
Due to their unusual properties they may also find applications in multiparty quantum communication tasks or quantum cryptography,
although not those based on Bell inequalities because we show here that up to three settings per party none of our states violate any Bell inequality.

The paper is organised as follows.
In the next section we review the construction of the $N$-qubit states without $N$-partite correlations, we show that one needs to search for a non-trivial generalisation of that method, and we introduce tools for dealing with higher dimensions.
In Sec.~\ref{SEC_RESULTS} we generalise the qubit construction and emphasise where it diverges from the original one.
We also prove that many of the states generated in this way are genuinely $N$-party entangled and provide the fraction of randomly generated input states for which the procedure produces ``entanglement without correlations''. 
Our final result is on possibility to violate a Bell inequality with these states.
We consider the state of three qutrits and using linear programming to find joint probability distribution reproducing quantum predictions~\cite{num}, 
we show that there is no Bell inequality with up to three measurement settings per party which could be violated.
We conclude in Sec.~\ref{SEC_CONC}.
 

\section{Review}
\label{SEC_REVIEW}


\subsection{No-correlation states of qubits}

An arbitrary state of $N$ qubits can be represented as:
\begin{equation}
\rho = \frac{1}{2^N} \sum_{\mu_1, \dots, \mu_N = 0}^3 T_{\mu_1 \dots \mu_N} \, \sigma_{\mu_1} \otimes \dots \otimes \sigma_{\mu_N},
\label{EQ_RHO_2}
\end{equation}
where $\{ \sigma_{0}, \sigma_{1}, \sigma_{2},\sigma_{3} \} = \{\hat 1,\sigma_x,\sigma_y,\sigma_z\}$ is the identity and Pauli matrices,
and coefficients $T_{\mu_1 \dots \mu_N} = \mathrm{Tr}(\rho \, \sigma_{\mu_1} \otimes \dots \otimes \sigma_{\mu_N})$ are so called correlation functions.
The correlation function is a standard statistical quantifier defined as the expectation value of a product of local measurement results.
If all $N$ observers are involved in measurements we talk about $N$-party correlations and the corresponding $T_{\mu_1 \dots \mu_N}$ has all the indices different than zero.
If some observers do not perform their measurements, it is sufficient to use a reduced density matrix to find correlations between them,
and one verifies that such lower order correlations are given by $T_{\mu_1 \dots \mu_N}$, where the indices corresponding to the observers who do not conduct measurements are set to zero.
For example, in a tripartite state correlation $T_{111}$ is tripartite and correlation $T_{110}$ is bipartite.

If the observers decide to measure general dichotomic observables, parameterised by vectors $\vec m_1, \dots, \vec m_N$,
then the resulting correlation function is related to the $T_{\mu_1 \dots \mu_N}$ coefficients by the tensor transformation law:
\begin{equation}
C(\vec m_1,\dots, \vec m_N) = \sum_{j_1,\dots,j_N = 1}^3 T_{j_1 \dots j_N} \, (\vec m_1)_{j_1} \dots (\vec m_N)_{j_N},
\end{equation}
where $(\vec m_n)_{j_n}$ is the component of the vector $\vec m_n$ along the $j_n$th axis.
In the present context this implies that it is sufficient to ensure that $T_{j_1 \dots j_N} = 0$ for all $j_1,\dots,j_N = 1,2,3$,
to guarantee that $N$-partite correlation functions vanish for arbitrary local measurements.

With this notation at hand Refs.~\cite{nc2,nc1} defined so called anti-state $\bar \rho$, to a given pure or mixed state $\rho$, by requiring that all $N$-partite correlations are opposite, i.e. $T_{j_1 \dots j_N}(\bar \rho) = - T_{j_1 \dots j_N}(\rho)$.
The following method was proposed to produce an anti-state to arbitrary input state of odd number of qubits, see also Refs.~\cite{es2015,es2018,guhne2018}.
By applying the map $\sigma_j \to - \sigma_j$ for $j = 1,2,3$, to every qubit, one notes that correlations between an odd number of observers are reversed.
This map is known as the universal-not gate~\cite{unot} and it is absent in the quantum formalism due to its anti-unitarity.
However, Ref.~\cite{nc1} demonstrates that its simultaneous application to every qubit produces a valid physical state.
In fact, by starting with a pure state one obtains in this way another pure state.
By evenly mixing a state with its anti-state
\begin{equation}
\rho_{\mathrm{nc}} = \frac{1}{2}(\rho + \bar \rho),
\label{ncstate}
\end{equation}
we therefore obtain $\rho_{\mathrm{nc}}$ with no correlations whatsoever between any odd number of particles.
In particular, applying this method to a system of odd-$N$ number of qubits produces a state with no $N$-partite correlations.

For future comparison let us also mention that in the case of even number of qubits there exists a strong numerical evidence that there is no anti-state to any genuinely multiparty entangled pure state \cite{nc2}.
As a consequence the construction of ``no-correlation'' states requires mixing of at least three pure quantum states. 
Indeed, families of genuinely multiparty entangled no-correlation states of even-$N$ number of qubits were provided in Ref.~\cite{nc2}.


\subsection{No straightforward generalisation}

Given the role universal-not gate plays in the qubit construction, one expects its extensions to higher dimensions will be useful when constructing no-correlation states of qudits.
For these systems, a natural generalisation of Pauli operators is given by the Heisenberg-Weyl operators
represented by $d \times d$ matrices:
\begin{eqnarray}
X&&=\left(\begin{array}{ccccc}
0&1&0&0&0\\
0&0&1&...&0\\
\vdots&\vdots&\vdots&\ddots&\vdots\\
0&0&0&\ldots&1\\
1&0&0&\ldots&0
\end{array}\right),\nonumber\\
Z&&={\rm Diag}(1,\omega_d,\omega_d^2,\ldots,\omega_d^{d-1}),\nonumber\\
\omega_d&&=\exp(i 2\pi / d).
\end{eqnarray}
One verifies that the set of operators $X^m Z^n$, with $m,n=0,1,\dots,d-1$, forms orthogonal basis with respect to the trace inner product.
The expected generalisation of the universal-not gate would then read:
\begin{eqnarray}
\mathcal{N}: X^m Z^n & \to & \omega_d^m \, X^m Z^n, \textrm{ for all } m,n=1,2,...,d-1\nonumber\\
Z^n & \to & \omega_d^n \, Z^n \textrm{ for all } n=1,2,...,d-1
\label{lambda}
\end{eqnarray}
so that evenly mixing $d$ states, the original one and the $d-1$ obtained from applying $\mathcal{N}^j$ (for $j = 1,\dots,d-1$) on every subsystem, 
would possess no $N$-party correlations whenever $N$ is not a multiple of $d$ (at least for prime $d$).

However, while the universal-not gate defined for qubits is a positive map which is not completely positive, the map $\mathcal{N}$ is not even positive in general. 
Take the simplest example of $|\psi\rangle=\frac{1}{\sqrt{2}}(|0\rangle+|1\rangle)$ in the qutrit domain, where the computational basis is the eigenbasis of operator $Z$.
One finds the following decomposition of this state in the Heisenberg-Weyl basis:
\begin{eqnarray}
|\psi \rangle \langle \psi | & = & \frac{1}{3}\Big( X^0 Z^0 + \frac{1}{2}(1 + \omega_3^2) X^0 Z^1 + \frac{1}{2}(1 + \omega_3) X^0 Z^2 \nonumber \\
& + & \frac{1}{2} X^1 Z^0 + \frac{1}{2} X^1 Z^1 + \frac{1}{2} X^1 Z^2 \nonumber \\
& + & \frac{1}{2} X^2 Z^0 + \frac{1}{2} \omega_3^2 X^2 Z^1 + \frac{1}{2} \omega_3 X^2 Z^2 \Big).   
\end{eqnarray}
After applying the map $\mathcal{N}$, i.e. multiplying the corresponding coefficients according to Eq.~(\ref{lambda}), we obtain the matrix:
\begin{equation}
\mathcal{N}(|\psi\rangle\langle\psi|)=\frac{1}{2} \left(\begin{array}{ccc}1&\omega_3&0\\\omega_3^2&0&0\\0&0&1\end{array}\right),
\end{equation}
which has negative eigenvalue $\frac{1}{4}(1- \sqrt{5})$.

Since we would like to have a general method, applicable to arbitrary input state, we should therefore resort to other procedures for finding the no-correlation states.
To this end we will utilise the generalised Gell-Mann matrices we now review.


\subsection{Generalised Gell-Mann basis}

The generalised Gell-Mann matrices in an arbitrary dimension $d$ constitute the generators of the Lie algebra associated to the special unitary group SU$(d)$. 
We denote $\lambda_{j,k}$ a matrix with $1$ on the $(j,k)$-th entry and 0 elsewhere.
The generalised Gell-Mann operators are represented by the following three groups of matrices:
\begin{eqnarray}
\textrm{Symmetric:} & & M^s_{j,k} = \lambda_{j,k}+\lambda_{k,j} \textrm{ for } 1 \leq j < k  \leq d \nonumber \\
\textrm{Antisymmetric:} & & M^a_{j,k} = -i(\lambda_{j,k}-\lambda_{k,j}) \textrm{ for } 1 \leq j < k  \leq d \nonumber \\
\textrm{Diagonal:} & & M^{g}_{j,k} = \sqrt{\frac{2}{l(l+1)}} \left( \sum_{i=1}^l\lambda_{j,j} - l \lambda_{l+1,l+1} \right) \nonumber \\
&& \textrm{ for } 1 \leq l \leq d-1.
\end{eqnarray}
They constitute the generators of SU$(d)$ group: $\mathcal{M} = \{M^s_{1,2}, M^s_{1,3}, ..., M^a_{1,2}, M^a_{1,3}, ..., M^{g}_{1}, ..., M^{g}_{d-1} \}$.
For simplicity, we re-index the elements of this set as follows $\mathcal{M} = \{M_{1}, M_{2}, ..., M_{d^2-1} \}$. 
The generalised Gell-Mann matrices are Hermitian $M_j = M_j^\dagger$, traceless $\textrm{Tr}(M_j)=0$, and orthogonal $\textrm{Tr}(M_iM_j)=2\delta_{ij}$, 
and can be used to decompose any quantum state in analogy to Eq.~(\ref{EQ_RHO_2}) for qubits:
\begin{equation}
\rho = \frac{1}{d^N} \sum_{\mu_1,\dots,\mu_N = 0}^{d-1} T_{\mu_1 \dots \mu_N} \, M_{\mu_1} \otimes \dots \otimes M_{\mu_N},
\end{equation}
where the $T_{\mu_1 \dots \mu_N}$ coefficients are the correlation functions, and $M_0$ denotes the identity.


\section{Results}
\label{SEC_RESULTS}

\subsection{The NOT map}

Let us define the following NOT map (generalisation of the universal-not for qubits) operating on a quantum system with dimension $d$:
\be\label{mapa}
\mathcal{N}_d(\cdot) := \sum_{a} \frac{M_a}{\sqrt{d-1}} \; (\cdot)^{\ast} \; \frac{M_a^\dagger}{\sqrt{d-1}},
\ee
where the sum is over all antisymmetric $d$-dimensional Gell-Mann matrices and complex conjugation is taken in the standard (computational) basis.
One readily verifies that:
\ben
\mathcal{N}_d(M_0) & = & M_0, \\
\mathcal{N}_d(M_j) &=& -\frac{1}{d-1}M_j, \textrm{ for } j \neq 0.
\label{EQ_NOT}
\een
We note the presence of the factor $\frac{1}{d-1}$ in the latter equation.
This factor is needed for the map to preserve positivity.
Indeed, as mentioned in Ref.~\cite{hvs}, for any pure state of a single qudit, with $d \ge 3$, the matrix obtained by replacement $T_{\mu} \to - T_{\mu}$ has negative eigenvalues.


\subsection{No-correlation states of qudits}

We apply the NOT map to each individual subsystem of a multipartite quantum state $\rho$.
The resulting state (the $d$-dimensional anti-state) is therefore:
\ben\label{Nn}
\bar{\rho} &=& (\mathcal{N}_d \otimes \dots \otimes \mathcal{N}_d) (\rho).
\een
By expanding $\rho$ in the generalised Gell-Mann basis and using (\ref{EQ_NOT}) we find
that $N$-partite correlations of the anti-state, $\bar T_{j_1 \dots j_N}$, are related to the correlations of the original state, $T_{j_1 \dots j_N}$, as follows:
\begin{equation}
\bar T_{j_1 \dots j_N} = \frac{(-1)^N}{(d-1)^N} T_{j_1 \dots j_N},
\end{equation}
where all the indices are not zero.
Accordingly, whenever $N$ is odd the anti-state has rescaled but opposite correlations to the original state.
Therefore they average out in the following uneven mixture:
\begin{eqnarray}
\rho_{\mathrm{nc}} & = & p \, \rho + (1-p) \, \bar \rho, \\
\textrm{with } p & = & \frac{1}{1+(d-1)^N}. \label{EQ_P}
\end{eqnarray}

Some comments are now in place.
First of all, since the NOT map is positive but not completely positive we should argue that $\bar \rho$ is a physical state, i.e. positive semi-definite matrix.
Indeed, the matrix $\rho^*$, obtained by complex conjugation in the standard basis, has the same eigenvalues as the original state.
Eq.~(\ref{mapa}) is then the Kraus representation of $\mathcal{N}_d$ with the Kraus operators $K_a = M_a / \sqrt{d-1}$. 
One directly verifies that for all $a$ we have $K_a^\dagger K_a \ge 0$ and $\sum_a K_a^\dagger K_a = \hat 1$, and hence $\mathcal{N}_d$ is a POVM (maps states to states).

Differently than in the case of qubits, if we start with a higher-dimensional pure state, its $d$-dimensional anti-state is in general mixed.
As a consequence the states with no correlations are of high rank, similarly to the case of even number of qubits.
Furthermore, the state $\rho_{\mathrm{nc}}$ in higher dimensions is guaranteed to have vanishing correlations only between all $N$ observers (for $N$ odd) whereas the qubit no-correlations states have vanishing correlation functions between any set of odd observers. 
As a final special feature of qubits we note that Eq.~(\ref{EQ_P}) reveals that for all higher dimensions the states without correlations are obtained by biased mixing with small contribution from the original state.

All this becomes particularly clear in the following low-dimensional examples.


\subsubsection{Example: three qutrits}

Consider an arbitrary state of three qutrits, i.e. each subsystem is of dimension $d=3$:
\ben
\rho &=& \frac{1}{27} M_0 \otimes M_0 \otimes M_0 \nonumber\\
&+& \frac{1}{18} \sum_{\pi_{i00}}\sum_{i=1}^8 T_{i00}\;M_i \otimes M_0 \otimes M_0  \nonumber\\
&+& \frac{1}{12} \sum_{\pi_{ij0}}\sum_{i,j=1}^8 T_{ij0}\;M_i \otimes M_j \otimes M_0 \nonumber\\
&+& \frac{1}{8} \sum_{i,j,k=1}^8 T_{ijk}\;M_i \otimes M_j \otimes M_k,
\een
where $M_0$ is $3\times3$ identity matrix, $\pi_{ijk}$ stands for the permutation of the indices $(i,j,k)$, and $T_{ijk} = \textrm{Tr}(\rho \; M_i\otimes M_j\otimes M_k)$ denotes the correlation functions.
Applying the NOT map to every subsystem gives the anti-state:
\ben
\bar{\rho} &=& \frac{1}{27} M_0 \otimes M_0 \otimes M_0 \nonumber\\
&-& \frac{1}{2} \frac{1}{18} \sum_{\pi_{i00}}\sum_{i=1}^8 T_{i00}\;M_i \otimes M_0 \otimes M_0  \nonumber\\
&+& \frac{1}{4} \frac{1}{12} \sum_{\pi_{ij0}}\sum_{i,j=1}^8 T_{ij0}\;M_i \otimes M_j \otimes M_0 \nonumber\\
&-& \frac{1}{8} \frac{1}{8} \sum_{i,j,k=1}^8 T_{ijk}\;M_i \otimes M_j \otimes M_k,
\een
where we explicitly kept the factors $\frac{(-1)^n}{(d-1)^n}$ in front of the $n$-partite correlations, for $n=1,2,3$.
By mixing $\rho$ and $\bar{\rho}$ with adequate proportions, $\frac19\rho + \frac89\bar{\rho}$, we obtain the state with no tripartite correlations:
\ben\label{stanpotr}
\rho_{\mathrm{nc}} &=& \frac{1}{27} M_0 \otimes M_0 \otimes M_0 \nonumber\\
&+&  \frac{1}{18} \sum_{\pi_{i00}}\sum_{i=1}^8 T'_{i00}\;M_i \otimes M_0 \otimes M_0  \nonumber\\
&+&  \frac{1}{12} \sum_{\pi_{ij0}}\sum_{i,j=1}^8 T'_{ij0}\;M_i \otimes M_j \otimes M_0,
\label{EQ_NC3}
\een
where the new correlation functions are given by $T'_{i00} = -\frac13 T_{i00}(\rho)$ and $T'_{ij0} = \frac13 T_{ij0}(\rho)$, respectively, and for all permutations of indexes.

\begin{figure}[!t]
	\centering
	\includegraphics[width=0.45\textwidth]{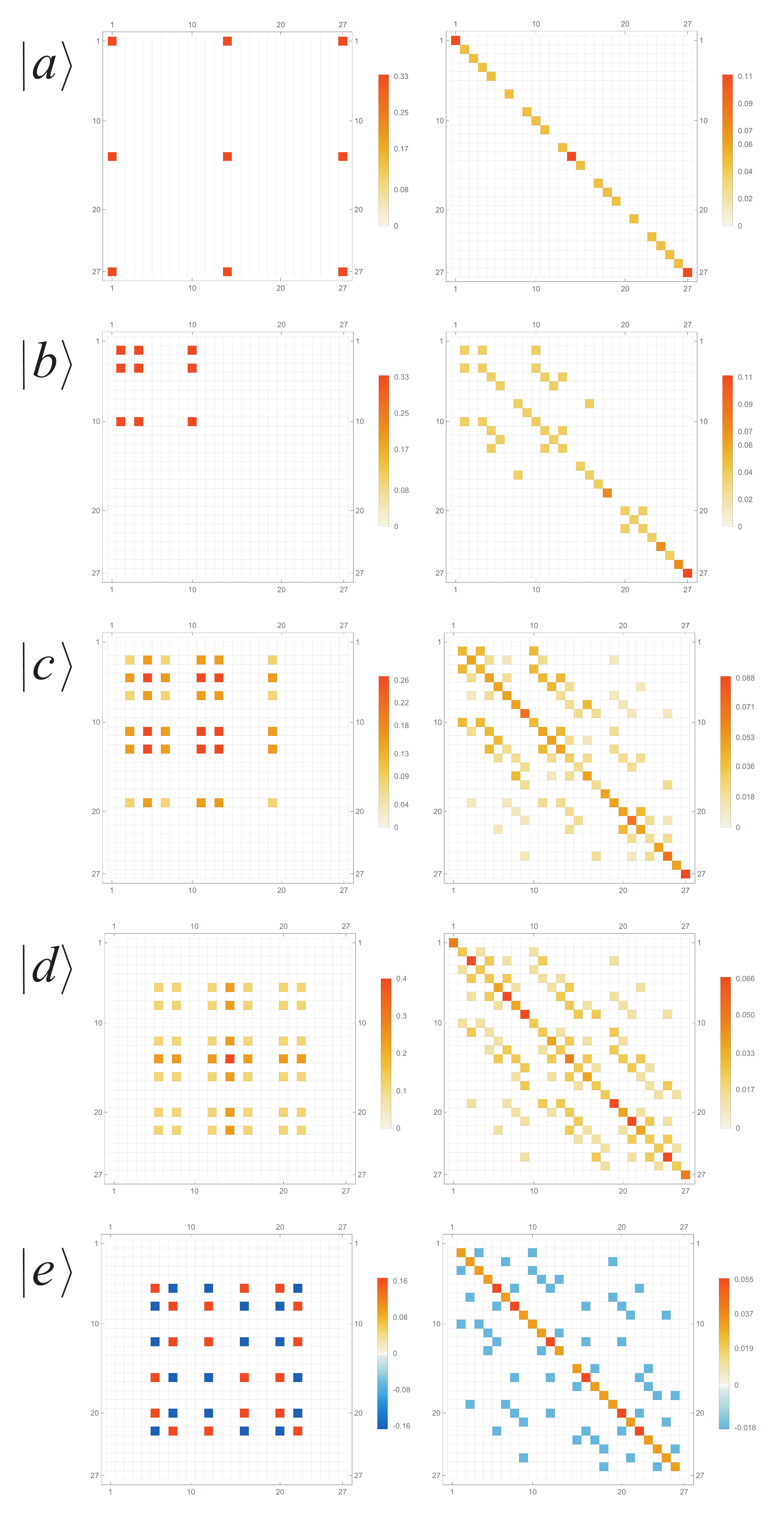}
	\caption{The left column presents density matrices (written in the standard / computational basis) of the states given in Eq.~(\ref{EQ_STATES}), in order they appear there.
	The right column presents density matrices of the corresponding states $\rho_{\mathrm{nc}}$ in Eq.~(\ref{EQ_NC3}). 
	(All discussed density matrices have only real elements.)
	The no-correlation states related to $|b\>$ and $|c\>$ are genuinely multiparty entangled, see Eq.~(\ref{EQ_WITNESS}).}
	\label{fig-states}
\end{figure}

We illustrate this method with the following states:
\ben
|a\> &=& \frac{1}{\sqrt3} (|000\>+|111\>+|222\>),  \nonumber \\
|b\> &=& \frac{1}{\sqrt3} (|001\>+|010\>+|100\>),  \nonumber \\
|c\> &=& \frac{1}{\sqrt{15}} (|002\>+|020\>+|200\> \nonumber \\
&+& 2(|011\>+|101\>+|110\>)),  \nonumber \\
|d\> &=& \frac{1}{\sqrt{10}}(|012\> + |021\> + |102\>  \nonumber \\
&+& |120\> +|201\> + |210\> + 2 |111\>), \nonumber \\
|e\> &=&\frac{1}{\sqrt{6}} (|012\> - |021\> - |102\>  \nonumber \\ 
&+& |120\> + |201\> - |210\>).
\label{EQ_STATES}
\een
The state $|a\>$ is a generalisation of the Greenberger-Horne-Zeilinger state to higher dimensions, see e.g.~\cite{huber},
states $|b\> , |c\>$ and $|d\>$ are the three qutrit Dicke states, 
and $|e\>$ is the state of vanishing total spin (singlet state) also known as the Aharonov state~\cite{adan}.
The density matrices of these states as well as their respective no-correlation states are presented graphically in Fig.~\ref{fig-states}.

In order to illustrate that states $\rho_{\mathrm{nc}}$ of qudits have non-vanishing lower-order correlations we calculate sum of all squared correlation functions between $n$ observers, for $n = 1,2,3$.
The results of these calculations, $\Sigma^n$, are presented in Tab.~\ref{tab1} both for the states in Eq.~(\ref{EQ_STATES}) and their corresponding no-correlation states.
We note though that the lower-order correlations present in $\rho_{\mathrm{nc}}$ are considerably suppressed compared to the original states.

\begin{table}
\center
	\begin{tabular}{lccc}
			\hline\noalign{\smallskip}
		$\rho $             & $\Sigma^1$ & $\Sigma^2$ & $\Sigma^3$ \\
				\noalign{\smallskip}\hline\noalign{\smallskip}
		$\rho^a$   & 0 & $\frac{8}{9}$ & $\frac{160}{27}$ \\
		$\rho^b$            & $\frac{4}{9}$ & $\frac{32}{27}$ & $\frac{128}{27}$   \\
		$\rho^c$            & $\frac{52}{225}$ & $\frac{704}{675}$ & $\frac{3584}{675}$   \\[0.5ex]
		$\rho^d$            & $\frac{16}{75}$ & $\frac{232}{225}$ & $\frac{3616}{675}$  \\[0.5ex]
		$\rho^e$     & 0 & $\frac{8}{9}$ & $\frac{160}{27}$  \\
			\noalign{\smallskip}\hline
	\end{tabular}
	\qquad
		\begin{tabular}{lccc}
		\hline\noalign{\smallskip}
		$\rho $             & $\Sigma^1$ & $\Sigma^2$ & $\Sigma^3$ \\
		\noalign{\smallskip}\hline\noalign{\smallskip}
		$\rho_{\mathrm{nc}}^a$   & 0 & $\frac{8}{81}$ & 0 \\
		$\rho_{\mathrm{nc}}^b$            & $\frac{4}{81}$ & $\frac{32}{243}$ & 0   \\[0.5ex]
		$\rho_{\mathrm{nc}}^c$            & $\frac{52}{2025}$ & $\frac{704}{6075}$ & 0   \\[0.5ex]
		$\rho_{\mathrm{nc}}^d$            & $\frac{16}{675}$ & $\frac{232}{2025}$ & 0  \\[0.5ex]
		$\rho_{\mathrm{nc}}^e$     & 0 & $\frac{8}{81}$ & 0  \\
			\noalign{\smallskip}\hline
	\end{tabular}	
	\caption{\label{tab1}Sums of all squared correlations between $n$ observers for various states.
	For example, $\Sigma^1 = \sum_\pi \sum_{j=1}^8 T_{j00}^2$, where $\pi$ denotes all permutations of indices $j00$.
	We compare original states presented in Eq.~(\ref{EQ_STATES}) with the corresponding no-correlation states.}
\end{table}


\subsection{Entanglement without correlations}

We have proposed a universal method which applies to an arbitrary input higher-dimensional multipartite state and generates states without $N$-partite correlations, for odd $N$.
Now we show that many states obtained in this way are genuinely $N$-party entangled.
We start with the states obtained from those in Eq.~(\ref{EQ_STATES}) and evaluate entanglement monotone $W$ proposed in Ref.~\cite{G2011} for the corresponding no-correlation states:
\begin{eqnarray}
W(\rho_{\mathrm{nc}}^a) & = & 0, \nonumber \\
W(\rho_{\mathrm{nc}}^b) & = & 0.0444,\nonumber \\
W(\rho_{\mathrm{nc}}^c) & = & 0.0147, \nonumber \\
W(\rho_{\mathrm{nc}}^d) & = & 0, \nonumber \\
W(\rho_{\mathrm{nc}}^e) & = & 0.
\label{EQ_WITNESS}
\end{eqnarray}
Positive value of the entanglement monotone indicates genuine multipartite entanglement and hence this proves it for the second and third state.

More generally, we have sampled uniformly at random, according to the Haar measure, $1000$ pure states of three qutrits (all of which turned out to be genuinely tripartite entangled)
and subjected them to our method of generation of no-correlation states.
The evaluation of the entanglement monotone $W$ on the resulting $3500$ no-correlation states proved tripartite entanglement in almost $27\%$ of them.


\subsection{Bell violation}

Bell inequalities have been constructed which revealed that many multi-qubit entangled states without correlations cannot be simulated with local hidden variable models~\cite{nc2,nc3,nc1}.
We were therefore searching, using the software introduced in Ref.~\cite{num}, for Bell inequalities which could be violated by the states introduced here.
Up to now this search proves that there exists no Bell inequality with up to three settings per party which is violated by any of the considered states $\rho^{a-e}_{\rm nc}$.


\section{Conclusions}
\label{SEC_CONC}

We introduced a method which applies to an arbitrary multipartite state of $N$ qudits and produces a state with no $N$-partite correlations, for odd $N$.
The correlation quantifier used here is the standard correlation function, i.e. the expectation value of the product of local measurement outcomes, and the observables are parameterised using the generalised Gell-Mann operators.
We stress that the states produced by our method show vanishing correlation functions between $N$ observers for arbitrary local measurements.
Yet, almost $27\%$ of the states generated starting from a random pure state of three qutrits are genuinely tripartite entangled. 
$N$-partite entanglement in such states is hence due to a combination of correlations between strictly less than $N$ particles.
The states put forward here will be a useful test-bed for candidate quantifiers of genuinely multiparty quantum and classical correlations satisfying natural postulates of Ref.~\cite{postulates}. 


\section*{Acknowledgments}

The work is supported by DFG (Germany) and NCN (Poland) within the joint funding initiative ``Beethoven2'' (2016/23/G/ST2/04273, 381445721). 
TP acknowledges the Singapore Ministry of Education Academic Research Fund Tier 2, Project No. MOE2015-T2-2-034.

\end{document}